\documentclass{ws-ijqi}

\begin{document} 

\markboth{Izumi Tsutsui}
{Physics of Points and Walls in Quantum Mechanics}

\catchline{}{}{}{}{}

\title{PHYSICS OF SINGULAR POINTS IN QUANTUM MECHANICS}

\author{IZUMI TSUTSUI${}^*$\,\, and \,\, TAM\'{A}S F\"{U}L\"{O}P${}^{\dagger}$}
\address{Institute of Particle and Nuclear Studies\\
KEK, Tsukuba 305-0801, Japan\\
${}^*$izumi.tsutsui@kek.jp, \,${}^{\dagger}$fulopt@post.kek.jp}
 
\maketitle

\begin{history}
\received{(21 November 2003)}
\end{history}

\begin{abstract}
Defects or junctions in materials serve as a source of interactions for particles, and in idealized limits they may be treated as 
singular points yielding contact interactions.  In quantum mechanics, these singularities accommodate an unexpectedly rich structure and thereby provide a variety of physical phenomena, especially if their properties are controlled properly.  Based on our recent studies, we present a brief review on the physical aspects of such quantum singularities
in one dimension.   Among the intriguing phenomena that the singularities admit, we mention strong vs weak duality, supersymmetry, quantum
anholonomy (Berry phase), and a copying process by anomalous caustics.  We also show that a partition wall as a singularity in a potential well can give rise to a quantum force which exhibits an interesting temperature behavior characteristic to the particle statistics. 

\end{abstract}


\section{Introduction}

If a system contains an object (or subsystem) different from its surroundings, and if the object is very small compared to the size of the system, then one may regard it as a \lq singularity\rq{}
in the system to a first approximation.  Such an object may be given by an isolated region forming a dot, or it may consist of a planar region forming a wall in the system.  In view of the negligible size of the object, we may study the physical property of the system by considering a point singularity interacting with particles such as electrons by contact interaction.  The outcome will furnish a basis for studying the physics of systems with singular objects given by quantum dots or
junctions in semiconductors, where the finite size effect may be considered as secondary to the zero size effect.  This analysis is also useful for analyzing systems from a long range point of view, where the singular object is reduced to a point (or a plane without thickness) in effect.

In classical mechanics, a singular 
point will
have no characteristics and is basically trivial.  In  
quantum mechanics, in contrast, it has been known that there are many, {\it distinct} singular points allowed, and if the system is one dimensional ({\it i.e.}, line) they form a $U(2)$ family~\cite{RS,AGHH}.  
The distinction among them lies in the connection conditions at the singularity  
obeyed by the wave functions, which can lead to entirely different physical consequences depending on the choice of the singularity.  Our interest then is to see how these singular points can be characterized mathematically, and to find what type of physical phenomena   
can be expected in systems with particles interacting with the $U(2)$ family of singularities if they are manipulated appropriately.   
The aim of the present paper is to provide a review of our investigations on these matters, 
and thereby to point out that, once a controllable singularity is introduced in the otherwise free system on a line, then
there appear many physically interesting properties, such as 
duality, anholonomy (Berry phase), supersymmetry and a copying process by quantum anomalous caustics.   We also mention a statistical 
aspect of systems with a singularity, which is the emergence of a force 
on a partition wall (regarded as singularity). 
The force is generated by distinct boundary conditions and exhibits an interesting dependence --- characteristic to the particle statistics (bosons or fermions) --- on the temperature and the particle number of the system.
Our consideration is restricted only to one dimensional systems, but the essential features of quantum singularities found here will also persist in higher dimensional systems. 

This paper is organized as follows.  After the Introduction, we briefly recall in section 2 how the $U(2)$ family of singularities appear in quantum mechanics on a line.  The physical meaning of the matrix notation used to characterize the family of singularities is then discussed in section 3.  Sections 4 and 5 are devoted to the physical phenomena afforded under the presence of the singularities mentioned above.  Section 6 discusses the quantum force on a partition wall and its temperature behavior.  Finally,  
we give our discussions in section 7.

\section{Quantum Description of a Singular Point on a Line}

Suppose that there is a point singularity, say, at
$x = 0$ on a line $-\infty < x < \infty$.  If the singularity is not accompanied by a potential, then
our Hamiltonian is the free one,
\begin{equation}
H =
-{{\hbar^2}\over{2m}}{{d^2}\over{dx^2}} ,
\label{eqn:ham}
\end{equation}

and the singularity will just impose a certain connection condition for wave functions at $x = 0$.  The connection condition arises from the requirement of the unitarity 
of the system,\footnote{Mathematically, this is equivalent to the self-adjointness of the Hamiltonian operator $H$ under the presence of the singularity.} 
which is ensured if the probability current
$
j(x) = - {{i\hbar}\over{2m}}\left(
(\psi^*)'\psi - \psi^* \psi' \right)(x)
$
is continuous at the singularity, 
$j(+0)=j(-0)$.  
This can be shown to be equivalent to the connection condition~\cite{AG,FT}
\begin{equation}
(U-I)\Psi+iL_0(U+I)\Psi'=0,
\label{eqn:bcon}
\end{equation}
where $U \in U(2)$ is called {\it characteristic matrix},
$L_0 \neq 0$ is a real (arbitrary) constant
and
\begin{equation}
\Psi =
\left(
\begin{array}{c}
{\psi (+0)}\\
{\psi (-0)}
\end{array}
\right),
\quad
\Psi' =
\left(
\begin{array}{c}
{\psi' (+0)}\\
{-\psi' (-0)}
\end{array}
\right),
\label{eqn:bvec}
\end{equation}
are vectors defined from the boundary values of the wave function
$\psi(\pm 0) = \lim_{x \to \pm 0} \psi(x)$ and their derivatives.
In other words, imposing the probability conservation as the sole requirement 
in quantum mechanics, we find the $U(2)$ 
freedom in specifying the connection 
condition, that is, there exist the $U(2)$ family of  
singularities possessing different connection conditions on a line.  For instance, if we choose $U = \sigma_1$ (where
$\sigma_i$ are Pauli matrices), the connection condition (\ref{eqn:bcon}) reduces
to
$\psi(+0) = \psi(-0)$,
$\psi'(+0) = \psi'(-0)$, which represents
the \lq free system\rq{} --- no actual
singularity at all.  On the other hand, the choice $U = -I$ gives the Dirichlet condition
$\psi(+0) = \psi(-0) = 0$ while $U = I$ gives the Neumann condition
$\psi'(+0) = \psi'(-0) = 0$.  

The last two choices provide two examples of 
connection conditions that prohibit the probability 
flow at the singularity.
All of these conditions represent, physically,  
an \lq infinite\rq{} partition wall with distinct characters. 
The general singularities that do not allow the probability flow are obtained by requiring $j(+0)=j(-0) = 0$, and the characteristic matrices that meet
this requirement are given by  
diagonal $U \in U(2)$.   
These form the so-called \lq separated subfamily\rq{} 
$U(1) \times U(1)$ in the $U(2)$ family, 
for which the connection condition (\ref{eqn:bcon})
reduces to the boundary condition,
\begin{equation}
\psi(+0)+L(\theta_+)\psi'(+0) = 0,
\qquad
\psi(-0)+L(\theta_-)\psi'(-0) = 0,
\label{eqn:redbc}
\end{equation}
where 
\begin{equation}
L(\theta_\pm) = L_0 \cot{{\theta_\pm}\over 2},
\label{eqn:spcprter}
\end{equation}
with $\theta_\pm$ being the phase parameters of ${\rm diag}\, U = (e^{i\theta_+}, e^{i\theta_-})$.  
The Dirichlet and the Neumann conditions are obtained, respectively, by choosing 
$L(\theta_\pm) = 0$ and $L(\theta_\pm) = \infty$.  In passing we note that, 
if the singularity is a wall (an end point of a positive half line), then the condition is simply $\psi(+0)+L(\theta_+)\psi'(+0) = 0$.  
An important point to be noted is that neither wave functions nor their derivatives are continuous at the singularity under a generic singularity (dot, partition or wall).

The forgoing argument applies almost unchanged to cases where  
the singularity arises as a divergent point of a potential, such as the Coulomb potential $V(x) = c/|x|$.  The only technical modification necessary is that the boundary vectors used in the connection condition (\ref{eqn:bcon}) must be slightly generalized as~\cite{TCF}
\begin{equation}
\Psi=
\left(
\begin{array}{c}
{W[\psi,\varphi_{1}]_{+0}}\\
{W[\psi,\varphi_{1}]_{-0}}
\end{array}
  \right),
\quad
\Psi'=
\left(
\begin{array}{c}
W[\psi,\varphi_{2}]_{+0}\\
- W[\psi,\varphi_{2}]_{-0}
\end{array}
  \right)
\label{eqn:gbvec}
\end{equation}
with the help of reference states $\varphi_{1}$,  $\varphi_{2}$ which are arbitrarily chosen to provide the self-dual
Wronskians
$
W[\phi,\psi](x) = \phi(x)\psi'(x)
-\psi(x)\phi'(x)
$.  
The modification is required because wave functions (and/or their derivatives), and hence the boundary vectors (\ref{eqn:bcon}), may diverge at the singularity under diverging potentials, while the Wronskians are always well-defined.   We note that (\ref{eqn:gbvec}) is a generalization of (\ref{eqn:bvec}), since by a suitable choice of the reference states one can regain (\ref{eqn:bvec}) from (\ref{eqn:gbvec}) when the states $\psi$ are well-defined at the singularity.

\section{Characteristic Matrix and the Spectral Space}

In order to investigate the physics implied by the singularity, we introduce the parametrization~\cite{TFC,CFT} of the characteristic matrix, 
\begin{equation}
U = V^{-1} D V,
\label{eqn:ude}
\end{equation}
with 
\begin{equation}
D = \left(
\begin{array}{cc}
e^{i\theta_+} & 0 \\
0 & e^{i\theta_-}
\end{array}
\right),
\qquad
V = e^{i{\mu\over 2}\sigma_2} e^{i{\nu\over 2}\sigma_3}
\label{eqn:udec}
\end{equation}
where $\theta_\pm \in [0, 2\pi)$ and
$\mu \in [0, \pi]$, $\nu \in [0, 2\pi)$.
The convenience of the parametrization may be recognized by the following observations. Notice, first, that any eigenstate $ \psi(x)$ with energy $E$ remains to be an eigenstate after the transformation of the parity ${\cal P}$ or the half-reflection ${\cal R}$ defined by
\begin{equation}
{\cal P}:
\psi \rightarrow
({\cal P}\psi)(x) := \psi( - x),
\qquad
{\cal R}: 
\psi \rightarrow
({\cal R}\psi)(x) := [\Theta(x) -
\Theta(-x)]\psi(x).
\label{eqn:dtrsf}
\end{equation}
The only effect caused by these can be found in the change of the connection
condition that the eigenstate obeys, and these are described by the corresponding change in the characteristic matrix,
\begin{equation}
U \buildrel {{\cal P} } \over
\longrightarrow  \sigma_1\, U\, \sigma_1, \qquad
U \buildrel {{\cal R} } \over
\longrightarrow  \sigma_3\, U\, \sigma_3.
\label{eqn:utrsf}
\end{equation} 
Obviously, the same can be observed with the transformation generated by the product ${\cal Q} = i{\cal P}{\cal R}$, which implies that this remains so even under any transformation given by a linear combination of the three generators $\{{\cal P} , {\cal Q}, {\cal R}\}$ which form an $su(2)$ algebra, as long as it does not change the norm of the state.   These general isospectral transformations induce  conjugations to the matrix $U$ by 
$U \rightarrow W^{-1} U W$ with $W \in SU(2)$.

\begin{figure}[t] 
\begin{center}
\includegraphics[width=0.8\linewidth]{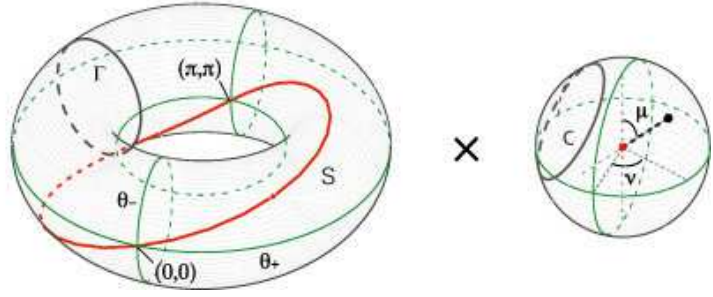}
\caption{$U(2)$ parameter space as a product of the torus $T^2$ determining the spectrum (the spectral space is its half $T^2/Z_2$ which is a M{\"o}bius strip) and the isospectral sphere $S^2$. }
\label{moebius}
\end{center}
\end{figure}

Thus we learn that the decomposition  (\ref{eqn:ude}) with (\ref{eqn:udec}) provides a split in the parameters into those that determine the sepctrum and those that do not, that is, $(\theta_+, \theta_-)$ and $(\mu, \nu)$.  The spaces of these parameters are, therefore, given by the spectral torus $T^2$ and the isospectral sphere $S^2$ (see Figure 1), respectively.  
More precisely, we shall see soon that the spectrum is unchanged under the interchange of the parameters $\theta_+ \leftrightarrow \theta_-$, and consequently the actual spectral space is given by $T^2/Z_2$ which is a M{\"o}bius strip with boundary~\cite{Moebius}.
  
The combinations $L(\theta_\pm)$ 
given in (\ref{eqn:spcprter}) set the scale of the system and turn out to be more useful than 
$\theta_\pm$.  {}For instance, for
$L(\theta_+) > 0$ and/or $L(\theta_-) > 0$, the singularity can support the bound states 
$\psi_\pm (x) \propto e^{- \vert x\vert/L(\theta_\pm)}$ where $L(\theta_\pm)$  represent the effective range of the particle trapped around the singularity.  
The isospectral parameters $(\mu, \nu)$, on the other hand, are related to the phase shift of the wave function at the singularity and the degree
of mixture of the limiting values of the wave function at $x = \pm 0$.  The 
characterization of the parameters discussed here hold true even under a symmetric potential, $V(-x) = V(x)$.

\section{Duality, Anholonomy and Supersymmetry}

Having furnished a formal basis to describe a generic quantum singularity on a line, we now turn to its physics.  Due to the nontrivial structure of the parameter space $U(2)$, various interesting phenomena can arise if we manipulate the parameters properly on the $U(2)$ space.  Here we mention three of them, duality, anholonomy and supersymmetry, which can be readily realized from what we have already.

\subsection{Duality}

The invariance of the spectrum under the interchange 
$\theta_+ \leftrightarrow \theta_-$ implies spectral duality for a 
pair of systems possessing singularities with the two parameters
interchanged.  
To see how this happens, for simplicity we restrict ourselves to parity
invariant singularities.  Note that, since the parity transformation ${\cal P}$ induces the change 
(\ref{eqn:utrsf}), parity invariant singularities are characterized by those $U$ satisfying
$\sigma_1\, U\, \sigma_1 = U$.  The general solution is given by
\begin{equation}
U = U(\theta_+, \theta_-)
  = e^{i(\theta_+P^+_1 + \theta_-P^-_1)},
\label{eqn:piu}
\end{equation}
with
$P^\pm_1 = {{1 \pm\sigma_1}\over 2}$, 
which is obtained by setting the ispospectral parameters
$(\mu, \nu) = (\pi/2, 0)$ in (\ref{eqn:udec}).

We also note that, to these parity invariant $U$ in (\ref{eqn:piu}) the half reflection
${\cal R}$ induces the exchange $\theta_+ \leftrightarrow
\theta_-$ through (\ref{eqn:utrsf}).  In view of the fact that the spectrum is preserved under ${\cal R}$, we realize that if the parameters $(\theta_+, \theta_-)$ of the two systems are the opposite of each other, then they share the same energy spectrum.  Moreover, since any eigenstate that arises under a parity invariant 
singularity can either be parity symmetric $\psi(-x) = \psi(x)$ or antisymmetric $\psi(-x) = -\psi(x)$, and since ${\cal R}$ swaps the parity, the corresponding eigenstates with the same energy in the two systems must have the opposite parity.  Note that ${\cal R}$ is an identity operation for the special type of singularities defined by $\theta_+ = \theta_-$, which are called \lq self-dual\rq.  It follows that in the self-dual case, which is indicated by the loop $S$ in Figure 1, the entire spectrum consists of doubly (or evenly) degenerate levels consisting of pairs of parity symmetric and antisymmetric states (see Figure 2).

\begin{figure}[t] 
\begin{center}
\includegraphics[width=0.7\linewidth]{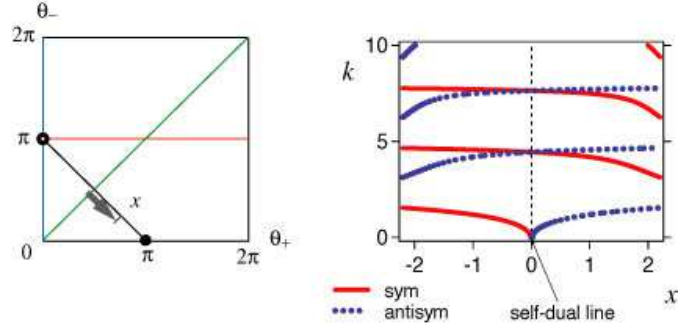}
\caption{Spectra of two systems with a singularity placed at the centre of an infinite well, where $k$ is the momentum with the energy $E = \hbar k^2/(2m)$.  The spectra,
drawn as a function of $x$ on the torus, are
identical for $x$ and $-x$, that is, for $(\theta_+, \theta_-)$ and $(\theta_-, \theta_+)$.}
\label{duality}
\end{center}
\end{figure}

Now we observe that, in this parity
invariant subfamily, the free system $U = \sigma_1$ arises 
at $(\theta_+, \theta_-) = (0, \pi)$.  This suggests
that we may consider \lq coupling constants\rq{} measuring the strengths of the
interaction at the singularity by
\begin{equation}
g_+(\theta_+) :=  \tan {\theta_+\over 2}\ ,
\qquad
g_-(\theta_-) := \cot {\theta_-\over 2}\ ,
\label{eqn:cconst}
\end{equation}
which vanish, $g_+(0) = g_-(\pi) = 0$, at the free point.  
In terms of these, we can interpret that the spectral duality holds for two systems with different 
coupling constants.   
In particular, if the parameters
fulfill
$\theta_+ = \theta_- \pm \pi$, then we find the reciprocal behavior
\begin{equation}
(g_+(\theta_+), g_-(\theta_-))
\buildrel {{\cal R} } \over
\longrightarrow
(- 1/g_+(\theta_+), -1/g_-(\theta_-)).
\label{eqn:cduality}
\end{equation}
In this case, the spectral duality can occur between two systems, one with a strong coupling and the other with a weak coupling.  This shows that the quantum singularity furnishes a simple example of {\it strong vs weak coupling duality}~\cite{TFC,CFT} which is normally discussed for more complicated systems such as (supersymmetric) gauge theory.

\subsection{Anholonomy}

Next we turn to the opposite situation where the spectral
parameters $(\theta_+, \theta_-)$ are fixed whereas
the isospectral parameters $(\mu, \nu)$ are free to vary.  
If we choose the free case $(\theta_+, \theta_-) = (0, \pi)$ for the fixed point, the collection of such singularities provides
the scale invariant subfamily, that is, they are invariant under the
Weyl scale transformation,
\begin{equation}
{\cal W}_\lambda: \quad
\psi(x) \longrightarrow
({\cal W}_\lambda\psi)(x) :=
\lambda^{1\over 2} \psi(\lambda x),
\label{eqn:weyltr}
\end{equation}
for real $\lambda$.  For our convenience, 
we consider a singularity of this kind placed  
at the centre of in an infinite well $[-l, l]$ and 
impose the Dirichlet boundary condition at the ends $x = \pm l$.  
Then the energy eigenstates are found to be
\begin{equation}
\psi_n(x)
= c_+(\mu)\, \xi_n^+(x)
+c_-(\mu) e^{i\nu}\, \xi_n^-(x),
\label{eqn:egenst}
\end{equation}
where 
\begin{equation}
c_\pm(\mu)
= \cos{\mu\over 2} \mp \sin{\mu\over 2},
\qquad
\xi_n^\pm(x)
= \sqrt{1\over l} \sin k_n(x\mp l) \Theta(\pm x) ,
\label{eqn:eeigenst}
\end{equation}
with 
$k_n 
= \left(n-{1\over 2}\right){\pi\over {2l}}$ for $n = 1,2,3,\ldots$.

Now, suppose that we have some means to control the isospectral parameters $(\mu, \nu)$ adiabatically.  Then we can consider a cyclic process of change along a loop $C$ on the
isospectral sphere
$S^2$ as shown in Figure 1.
After completing the cycle, each eigenstate returns to the initial
one modulo a phase pertinent to the state, $\psi_n \rightarrow e^{i\gamma(C)}\psi_n$.  This is the {\it phase anholonomy} (or the Berry phase) and is evaluated to be
\begin{equation}
\gamma(C) 
= \oint_C A, \qquad
A
= i\langle \psi_n \vert
d \psi_n \rangle
=  -{1\over 2}(1+\sin\mu)\, d\nu,
\label{eqn:bcn}
\end{equation}
where $d$ is the exterior derivative in the parameter space~\cite{CFT}.
Note that the curvature $F = dA$ is just the magnetic field of the Dirac monopole,
$
F = -{1\over 2}\cos\mu\,d\mu\,d\nu
$.

\begin{figure}[t] 
\begin{center}
\includegraphics[width=0.7\linewidth]{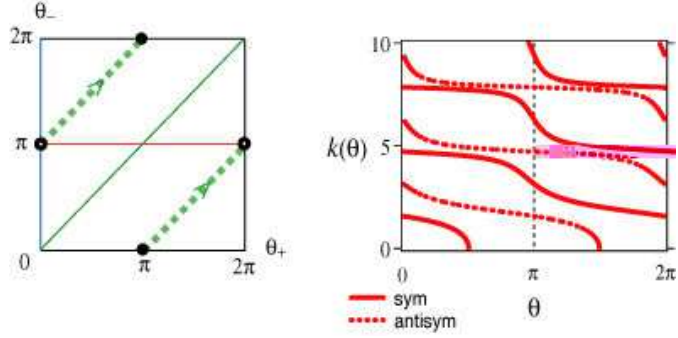}
\caption{Spectral change along the cyclic process paralleling the self-dual loop $S$ distanced by $\pi$.  Both symmetric and antisymmetric levels turn to be a lower (or higher depending on the direction) level by two after the cycle, even though the whole spectrum is unchanged. }
\label{loops}
\end{center}
\end{figure}

A similar adiabatic cyclic process may also be considered on the spectral torus,
instead of the isospectral sphere.  For instance, we may change the spectral parameters along a loop which winds over the surface of the torus nontrivially (the simplest will be the loop $\Gamma$ in Figure 1).  After
completing one cycle, we find a different type of anholonomy, where each level does not return to the initial one, even though the entire spectrum as a whole is recovered.  The response of the spectral change depends on the cycle one chooses; for example, if the loop is taken to be the one paralleling the self-dual loop $S$ with distance $\pi$ as shown in Figure 3, then 
the discrete momenta $k$ corresponding to symmetric states are given by $k = k(\theta_+)$ while those corresponding to antisymmetric states are
$k = k(\theta_- \pm \pi)$, where the function $k(\theta)$ is determined by 
\begin{equation}
k(\theta)\, L_0 \cot k(\theta)l = \tan {{\theta}\over 2}.
\label{eqn:nahsc}
\end{equation}

The resultant spectral change in Figure 3 shows the {\it level anholonomy}, as they shift by two after one cycle~\cite{CFT}.  This double spiral structure of energy levels along the loop may in future be used to implement a specific physical process like the one considered for the holonomic quantum computation in systems exhibiting the Berry phase.

\subsection{Supersymmetry}

The double degeneracy occurring under self-dual singularities
suggests that these systems may accommodate
supersymmetry (SUSY).  In fact, one can show that for a certain
class of systems, including those with a special type of self-dual singularities, it is possible to associate SUSY without necessarily yielding degeneracy in the spectrum.

To see this, let us rewrite our system into a
set of two systems each of which defined on a half line.  
There, we employ, instead of the wave functions $\psi(x)$ and the
Hamiltonian $H$ in (\ref{eqn:ham}), the two-component wave functions and the corresponding Hamiltonian
\begin{equation}
\Psi(x) =  \left(
\begin{array}{c}
{\psi_+(x)}\\
{\psi_-(x)}
\end{array}
\right),
\qquad
H = -{\hbar^2\over {2m}}\frac{d^2}{dx^2}\otimes I,
\label{eqn:wfham}
\end{equation}
where 
$\psi_\pm(x) := \psi(\pm x)$ for $x > 0$
and $I$ is the $2 \times 2$ identity matrix.
Our supercharge is assumed to take the form
\begin{equation}
Q =
-i\lambda\frac{d}{dx}
\otimes\sigma_{\vec{a}}+
{\bf 1} \otimes
\sigma_{\vec{b}},
\label{eqn:sscharge}
\end{equation}
where 
$\lambda = \hbar/2\sqrt{m}$ 
and
\begin{equation}
\sigma_{\vec{a}} = \sum_{i = 1}^3{a_i \sigma_i},
\quad
\sigma_{\vec{b}} = \sum_{i = 1}^3{b_i \sigma_i},
\quad
\vert \vec{a}\vert = 1 ,
\quad
\vec{a}\cdot \vec{b} = 0,
\label{eqn:ssche}
\end{equation}
with real vectors $\vec{a}$, $\vec{b}$.  The conditions (\ref{eqn:ssche})
ensure the formal relation $2 Q^2 = H+\vert {\vec b}
\vert^2$.  Thus, if we absorb the constant $\vert {\vec b} \vert^2$ into
the Hamiltonian (which causes only the corresponding constant energy shift),
we obtain, for a set of independent supercharges $Q_i$ for $i = 1,\ldots,N$ which are normalized properly, the standard SUSY
algebra,
\begin{equation}
\{Q_i, Q_j\} = H\, \delta_{ij}.
\label{eqn:ssalg}
\end{equation}

The important question we need to address at this point is whether the
supercharge $Q$ leaves the given connection condition invariant, at least for
energy eigenstates, because otherwise the SUSY transformation acts on states allowed under different singularities and hence the SUSY is not defined within a single system.  
Thus our demand for SUSY to exist is that, given a singularity specified by $U$, both the state $\Psi(x)$ and $Q \Psi(x)$ fulfill the same connection condition.
To answer this question, 
we first note that, if the state
$\Psi(x)$ fulfills the connection condition (\ref{eqn:bcon}), then for any $W \in U(2)$
the state $W \Psi(x)$ fulfills the same connection condition with $U$ replaced by $WUW^{-1}$.  This implies that,
if the pair $(U, Q)$ satisfies the above demand, so does the
pair $(WUW^{-1}, WQW^{-1})$.  Note also that $WQW^{-1}$ is again in the form (\ref{eqn:sscharge}), and hence by 
choosing in particular
$W = V$ with $V$ appearing in the decomposition (\ref{eqn:ude}), we find that the pair $(D, VQV^{-1})$ also satisfies the demand.  For this
reason, with no loss of generality, we may assume that  
$U$ is diagonal.  We then find, by a straightforward inspection, that  
the required condition is fulfilled if $\theta_+ = \theta \neq 0$ and
$\theta_- = \pi$ (and vice versa), and further if the supercharge takes the form
\begin{equation}
Q = V^{-1} \,q(\alpha, c; \theta)\, V,
\label{eqn:spcg}
\end{equation}
with
\begin{equation}
q(\alpha, c; \theta)
={} -i\lambda\frac{d}{dx}\otimes
e^{-i{\alpha\over 2}\sigma_3}\sigma_1 e^{i{\alpha\over 2}\sigma_3} 
{}+ {\bf 1}
\otimes
\left[-{\lambda\over{L(\theta)}}
e^{-i{\alpha\over 2}\sigma_3}\sigma_2
e^{i{\alpha\over 2}\sigma_3}
+ c\,
\sigma_3\right],
\label{eqn:spss}
\end{equation}
where $L(\theta)$ is the scale parameter defined in (\ref{eqn:spcprter}).
Since $\alpha$ is arbitrary, there are two independent
supercharges, {i.e.,} the system has an $N = 2$ SUSY~\cite{UTone,UTtwo} as long as one of the two eigenvalues of the characteristic matrix $U$ is $-1$ and the other not $-1$.

\begin{figure}[t] 
\begin{center}
\includegraphics[width=0.5\linewidth]{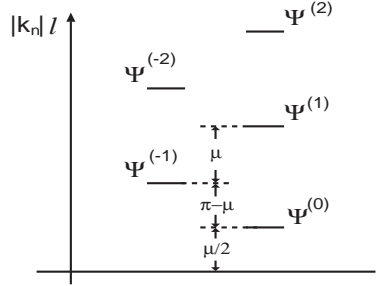}
\caption{Energy levels of the $N = 1$ SUSY system occurring
under the singularity with $U = V^{-1} \sigma_3 V$.  The levels, which are dependent on $\mu$ because the boundary conditions break the parity invariance, 
are not degenerate unless $\mu = 0$ or $\pi$.}
\label{susy}
\end{center}
\end{figure}

When we put the singularity in an infinite well, then the SUSY may be found depending on the boundary conditions imposed at the ends.  Various combinations, and accordingly various types of SUSY systems arise, and before we proceed we mention one of these.  Consider the boundary condition
\begin{equation}
\psi^\prime_+(l) = 0, 
\qquad 
\psi_-(l) = 0,
\label{eqn:ndbc}
\end{equation}
and the singularity specified by $U = V^{-1} \sigma_3 V$ which gives the connection condition
\begin{equation}
e^{i\nu} \psi_+(+0) - \cot {\mu\over 2}\, \psi_-(+0) = 0, 
\qquad 
e^{i\nu} \psi_+^\prime(+0) + \tan {\mu\over 2}\, \psi_-^\prime(+0) = 0.
\label{eqn:cbtwo}
\end{equation}
Then, the system admits an $N = 1$ SUSY with the supercharge $Q$
for $\vec{b} = 0$ and the energy eigenstates
\begin{equation}
\Psi^{(n)}(x) = N^{(n)}	 
		\left( \matrix{ - e^{-i\nu} \cos k_n (x-l) \cr
        \sin k_n (x-l) } \right) ,
\qquad
k_n = {{n\pi + \mu/2}\over{l}},
\label{eqn:eeigen}
\end{equation}
for $n \in Z$.  Each eigenstate is invariant
under the SUSY transformation generated by $Q$, and
the energy levels $E^{(n)} = \hbar^2 k_n^2/(2m)$ 
are not degenerate unless $\mu = 0$ or $\pi$ (see Figure 4).  
Supersymmetry, therefore, does not necessarily imply degeneracy in levels.

\section{Quantum Tunneling and Copy}

When the singularity is accompanied with a potential $V(x)$, we can expect various
interesting phenomena by combining the property of the quantum singularity
and the property pertinent to the potential.  One such example is provided by \begin{equation}
V(x) = \frac{m {\omega^2}}{2} {x^2} +
g\frac{1}{x^2},
\label{eqn:cstpot}
\end{equation}

which is known to admit \lq caustics\rq~\cite{Schulman}, a phenomena that arises when the classical dynamics of the system exhibits a certain type of singularity in the initial value problem, with the typical example being the periodic recurrence of the harmonic oscillator.  In fact, for $g > 0$ the dynamics of the system resembles the harmonic oscillator, and the only essential difference is that the recurrence occurs in each of the half lines because the system splits into two subsystems at 
the singularity $x = 0$ due to the infinite potential wall there.

In quantum mechanics, the situation is quite different.  
To see this, note first that the general solution
for the Schr{\"o}dinger equation
$H {\psi_n}(x) = {E_n}{\psi_n}(x)$ (where now $H$ has the potential term $V(x)$)
is given by a linear combination of
the two independent solutions,
\begin{equation}
\phi^{(1)}_n(x)
:={y^{c_1 - 1/2}}e^{-{y^2}/{2}}
F\left(\frac{{c_1}-\lambda_n}2,{c_1};{y^2}\right), 
\label{eqn:indgensla}
\end{equation}
and

\begin{equation}
{\phi^{(2)}_n}(x)
:={y^{c_2 - 1/2}}e^{-{y^2}/{2}}
F\left(\frac{{c_2}-\lambda_n}2,{c_2};{y^2}\right),
\label{eqn:indgenslb}
\end{equation}
where $F(\alpha,\gamma;z)$ is the confluent hypergeometric
function, $\lambda_n = E_n/\hbar\omega$ and 
\begin{equation}
c_1 = 1 + a, 
\qquad
c_2 = 1 - a, 
\qquad
a=
{1\over 2}\sqrt{1+\frac{8mg}{\hbar^2}},
\qquad
y =\sqrt{\frac{m\omega}\hbar} \, x.
\label{eqn:solprmt}
\end{equation}
The point is that, if the coupling constant $g$ is in the range
\begin{equation}
0 < g < {{3\hbar^2}\over{8m}},
\label{eqn:ccrange}
\end{equation}
we have ${1\over 2} < a < 1$, and therefore both of the two solutions
(\ref{eqn:indgensla}), (\ref{eqn:indgenslb}) are square integrable, even though $\phi^{(2)}_n$
may diverge at
$x = 0$.  The existence of the solution which does not vanish (actually diverges) at $x = 0$ implies that in quantum mechanics the system does {\it not} split there, in contrast to the classical case.

The general solution $\psi_n (x)$ is then given by a linear
combination of these two solutions with arbitrary coefficients
$N_{\rm R}^{(s)}$ and $N_{\rm L}^{(s)}$ for $s = 1$, 2, which can differ on the positive and negative sides,
\begin{eqnarray}
\psi_n (x)
&=&
[{N_{\rm R}^{(1)}}\phi^{(1)}_n(|x|)
+{N_{\rm R}^{(2)}}\phi^{(2)}_n(|x|)]\Theta(x) \nonumber \\
&{}&\quad
+[{N_{\rm L}^{(1)}}\phi^{(1)}_n(|x|)+{N_{\rm
L}^{(2)}}\phi^{(2)}_n(|x|)]\Theta(-x),
\label{eqn:solgen}
\end{eqnarray}
where 
$\Theta(x)$ is the Heaviside step function.
Let us now choose 
the reference modes
\begin{equation}
\varphi_{1}(x) := \sqrt{\frac{\hbar}{m\omega}}\,
\phi_{n_0}^{(1)}(\vert x\vert)\left[\Theta(x) -
\Theta(-x)\right], 
\qquad
\varphi_{2}(x) := 
\frac{1}{c_2  -  c_1}\,
\phi_{n_0}^{(2)}(\vert x\vert),
\label{eqn:rrmode}
\end{equation}
which are the solutions in (\ref{eqn:indgensla}), (\ref{eqn:indgenslb})
with $n = n_0$ for which $\lambda_{n_0} = 0$.  Using these
in (\ref{eqn:gbvec}) to get the boundary vectors,
\begin{equation}
\Psi= ({c_1}-{c_2})
\pmatrix{
N_{\rm R}^{(2)}\cr
N_{\rm L}^{(2)}
},
\qquad
\Psi'=\sqrt{\frac{m\omega}\hbar} \, 
\pmatrix{
N_{\rm R}^{(1)}\cr
N_{\rm L}^{(1)}
},
\label{eqn:bvcaust}
\end{equation}
and then plugging these in the connection condition
(\ref{eqn:bcon}), 
one obtains the spectral condition,
\begin{equation}
\frac{1}{{c_1}-{c_2}}
\sqrt{\frac{m\omega}\hbar}\frac{\Gamma\left(({c_1}-\lambda_n)/2\right)}
{\Gamma\left(({c_2}-\lambda_n)/2\right)}
\frac{\Gamma(c_2 )}{\Gamma(c_1)} = \frac{1}{L(\theta_\pm)}.
\label{eqn:ccra}
\end{equation}
Thus one finds that, in general, there exist two series of energy levels, one
specified by $L(\theta_+)$ and the other by $L(\theta_-)$.
{}For instance, if the singularity is free, $U = \sigma_1$, then we have
the two series of eigenstates,
\begin{equation}
\psi_n^{(1)}(x)
=
N^{(1)}\, \phi^{(1)}_n(\vert x\vert)
\left[\Theta(x) - \Theta(-x)\right], \qquad
\psi_n^{(2)}(x)
=
N^{(2)}\, \phi^{(2)}_n(\vert x\vert) ,
\label{eqn:indgen}
\end{equation}
with the eigenvalues,
\begin{equation}
E^{(1)}_n =(2n+{1 + a})\hbar\omega, \qquad
E^{(2)}_n =(2n+{1 - a})\hbar\omega,
\label{eqn:elevels}
\end{equation}
for $n = 0, 1, \ldots$.  In the limit $g \to 0$ ($a \to 1/2$)
these states reduce to the familiar eigenstates of the harmonic oscillator as
expected.  Such a smooth limit does not exist, however, for other
singularities, {\it e.g.,} at the Dirichlet point $U = -I$, one obtains
the doubly degenerate energy levels
$E_n = (2n+{c_1})\hbar\omega$ which do not reduce to those of the harmonic oscillator.  This case $U = -I$ corresponds to the conventional connection condition used to
provide the solutions in the Calogero model~\cite{Calogero}.

Having solved the quantum system, we now see that the singularity allows quantum tunneling. Indeed, for the free case $U = \sigma_1$, for instance, the generic state
\begin{equation}
\psi(x) = \sum_n(c_n^{(1)}\psi_n^{(1)}(x) +
c_n^{(2)}\psi_n^{(2)}(x))
\label{eqn:genstate}
\end{equation}
has the probability current at the singularity
\begin{equation}
j(\pm 0) =
\frac{ia\hbar}{m}\sum_{n, l}
\left\{ (c_n^{(1)})^*c_l^{(2)} - (c_n^{(2)})^*c_l^{(1)}\right\},
\label{eqn:pcrnt}
\end{equation}
which is non-vanishing.  The tunneling is seen generically, except for 
those singularity belonging to the separated subfamily mentioned earlier.
Another evidence may be gained from the transition amplitude,
\begin{equation}
K({x_f},{t_f};{x_i},{t_i})
= \langle x_f \vert e^{-{i\over\hbar} H (t_f - t_i)} \vert x_i \rangle,
\label{eqn:tramp}
\end{equation}
which can be evaluated exactly with the help of the solutions obtained above.
We then find that, for the transition time $T := t_f - t_i \neq k\pi/\omega$
($k = 0, 1, 2,\ldots$), the amplitude is expressed in terms of the modified Bessel function,
and from it we learn that the transition across the singularity is indeed allowed.

\begin{figure}[t] 
\begin{center}
\includegraphics[width=0.67\linewidth]{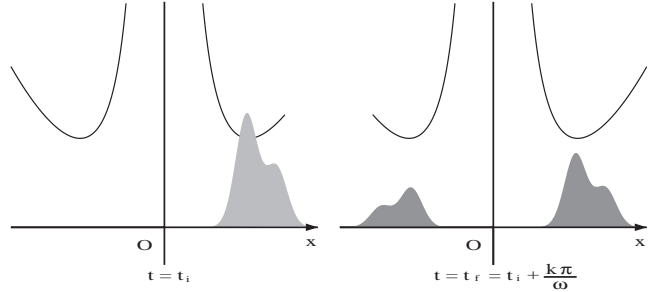}
\caption{Process of quantum copy through the caustics anomaly.  
At every period $T =
k\pi/\omega$, a mirror image of the original profile on $x > 0$ 
emerges on the
other side $x < 0$.  The relative 
size of the mirror image depends on $a$ and
$k$.}
\label{qcopy}
\end{center}
\end{figure}

The remarkable point is that, at the periods of oscillation
$T=k\pi/\omega$, the amplitude turns out to be
\begin{equation}
K({x_f},{t_f};{x_i},{t_i})
=
(-1)^k \cos(ak\pi)\delta({x_f}-{x_i})
+ i(-1)^k \sin(ak\pi)\delta({x_f}+{x_i}).
\label{eqn:trptd}
\end{equation}
The first term on the r.h.s.~corresponds to the return of
the particle to its initial position (which is the classical caustics), while the second term corresponds to the tunneling of the particle which reaches
the mirror point of the initial position with respect to the singular wall.  This shows that the classical caustics
has been modified at the quantum level ({\it i.e., caustics anomaly}), in such a way that we can now have the mirror image of the original
profile prepared at the initial time $t = t_i$, with the weight factors being the functions of the parameter $a$ determined from the coupling constant $g$ (and the characteristic matrix $U$ for the general case).  In other words, one can \lq copy\rq{} an original
profile prepared on the $x > 0$ side to the other $x < 0$ side after
the periods, and that this can be done with desirable weight factors if one can control the
relevant parameters of the factors freely~\cite{MT}.  Note that this copying process is not in conflict
with the no-go theorem~\cite{WoottersZurek} of quantum cloning, because the process takes
place in one Hilbert space rather than two as presumed in the theorem.

\section{Quantum Force on a Partition Wall}

Let us finish our discussion with another example to exhibit how
remarkably distinct physics arises for distinct possible dots and walls.

Consider an interval $[-l, l]$ bordered by Dirichlet
reflecting walls, $\psi(\pm l) = 0$. Suppose we insert a separating dot at
the centre with $L(\theta_{+}) = \infty$, $L(\theta_{-}) = 0$, in other
words, a partition wall that imposes the Neumann condition from the right
and the Dirichlet one from the left.  Suppose also that we put $N$ identical bosonic
particles into each of the two half wells, which we keep at the same
temperature, and calculate the quantum statistical average forces (or
pressure) acting on the partition from the right and the left. Notably,
the only difference between the circumstances on the two half wells is
the distinct reflecting property of the separating wall from the two
directions. We will see
that, due solely to this fact, the net force will be nonvanishing and
reaching arbitrarily large values at high enough temperatures~\cite{FMT}.

To see this, we recall first that 
the right and left half wells admit the energy levels
$E^{\pm}_n = e^{\pm}_n {\cal E}$, $n = 1$, $2$, $3$ $\ldots$, with
  \begin{equation}
e^{+}_n = \left(n-{1\over 2}\right)^2, \qquad
e^{-}_n = n^2, \qquad 
{\cal E} = {{\hbar^2}\over{2m}}\left({{\pi}\over{l}}\right)^2,
  \label{unitenergy}\end{equation}
where we use hereafter the indices \lq $+$\rq{} and \lq $-$\rq{} to indicate quantities for the right and left half wells, respectively.
The particles will distribute among these eigenstates according to the
Bose-Einstein statistics,
  \begin{equation}
N_n^\pm = {1\over{e^{\alpha^\pm + e^\pm_n / t} - 1}},
  \label{popl} \end{equation}
where $t$ is the dimensionless temperature parameter $t = k T / {\cal E}$,
and the temperature dependent $\alpha^\pm = \alpha^\pm(t)$ are determined by $N =
\sum_{n} N_n^\pm$. The forces acting from the right and the left are
given by
  \begin{equation}
F^\pm = - \sum_n {{\partial E_n^\pm}\over{\partial l}} N_n^\pm =
\frac{2 {\cal E}}{l} \sum_n e^\pm_n N^\pm_n. 
  \label{nfp} \end{equation}

\begin{figure}[t] 
\begin{center}
\includegraphics[width=0.5\linewidth]{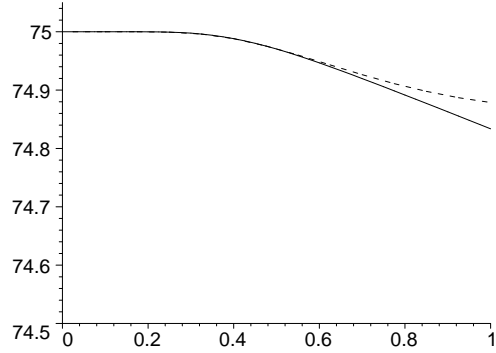}
\caption{The dimensionless net force $ \frac{l}{2 {\cal E}} \Delta F(t) $
for $N = 100$, in the temperature region $t < 1$, obtained by a numerical
computation (solid line), and approximated by (\ref{nfres}) (dashed line).}
\label{fig:low} 
\end{center} 
\end{figure}

For low temperatures, most of the particles are in the ground state, 
and $N^\pm_n$ decrease exponentially fast for higher $n$.  Consequently, the net force
will be contributed essentially by the ground and first excited levels (see Figure~\ref{fig:low}),
giving
  \begin{equation}
\Delta F(t) \approx \frac{2 {\cal E}}{l}
\left[ \frac{3}{4} N + 3 \, e^{-3/t} - 2 \, e^{-2/t} \right]
  \label{nfres} \end{equation}

\begin{figure}[t] 
\begin{center}
\hskip -0.ex\includegraphics[width=0.45\linewidth]{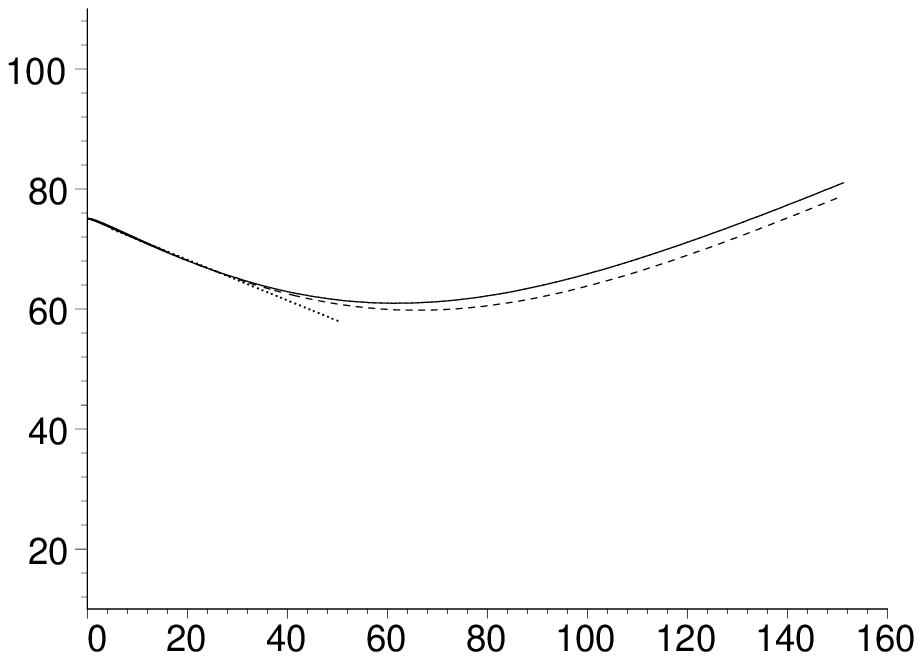}\hskip 7.ex%
\includegraphics[width=0.45\linewidth]{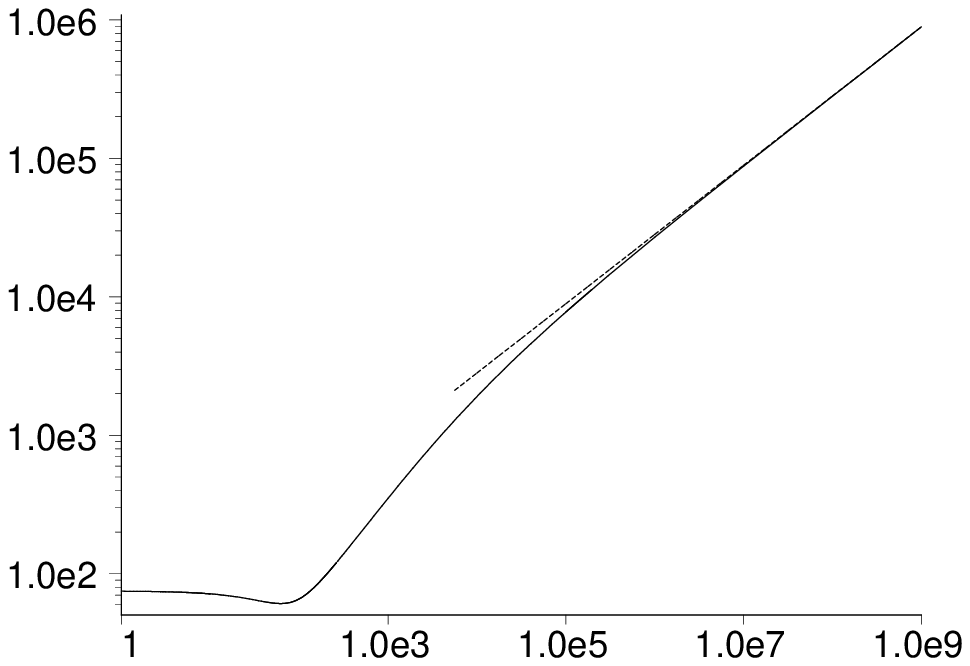} 
\caption{%
Left: The
dimensionless net force $ \frac{l}{2 {\cal E}} \Delta F(t) $ for
$N = 100$, in the temperature region $0 < t < 160$, obtained by a
numerical computation (solid line), and approximated by (\ref{nflt})
(dotted line), and by (\ref{qnf}) using (\ref{aat}) (dashed line).
Right: The dimensionless net force $ \frac{l}{2 {\cal E}} \Delta
F(t) $ for $N = 100$, obtained by a numerical computation (solid
line), and approximated for high temperatures with (\ref{aaj})
(dashed line). The figure is double logarithmic.}
\label{fig:med} 
\end{center} 
\end{figure}

As temperature is increased, more and more levels enter, and the
net force starts to decrease approximately linearly. This behavior
can be accounted for by using a heuristic argument \cite{FMT} which 
classifies the energy levels into three classes and estimates
the contribution of each class in turn.  The result,
  \begin{equation}
\Delta F(t) \approx \frac{2 {\cal E}}{l}
\left[ {3\over 4} N - {t\over{(e - 1)^2}} \right]
  \label{nflt} \end{equation}
proves to be satisfactory up to $t \approx 2N/3$, where the net
force reaches a minimum and starts to increase afterwards (see
Figure~\ref{fig:med}).

To explain this minimum and the increase following it with an
analytic approximation formula, let us replace the infinite sums (\ref{nfp}) and $N = \sum_{n}
N_n^\pm$ with corresponding integrals, which is allowed
in this temperature region.  Assuming $|\alpha^\pm| < 1$
as well, one then obtains \cite{FMT}
  \begin{equation}
\Delta F \approx \frac{2 {\cal E}}{l} \left[ \left( N t + \frac{35}{96}
\sqrt{\pi} t^{3/2} \right) \left( \alpha^+ - \alpha^- \right)
+ \left( \sqrt{ e_1^+ } - \sqrt{ e_1^- } \right) t \right] ,
  \label{qnf} \end{equation}
where the $\alpha^\pm$ are to be determined from
  \begin{eqnarray}
N & \approx & \frac{1}{ \alpha^\pm + e^\pm_1 / t }
+ \frac{1/2}{ \alpha^\pm + e^\pm_2 / t } - \frac{3}{4}
- \frac{ \sqrt{( 2 - \alpha^\pm ) t} - \sqrt{e^\pm_2} }{2} \nonumber \\
& & + \sqrt{ \frac{t}{|\alpha^\pm|} } \left[ A \left(
\sqrt{ \frac{ |\alpha^\pm| t }{e^\pm_2} } \right)
- A \left( \sqrt{ \frac{ |\alpha^\pm| }{ 2 - \alpha^\pm } } \right) \right],
  \label{aat} \end{eqnarray}
with $A$ denoting the arctan function for positive $\alpha^\pm$ and the
arctanh function for negative $\alpha^\pm$. It is not easy to express the
solution $\alpha^\pm$ of (\ref{aat}) directly via an approximate analytic
formula, but by calculating the solution numerically and
applying it in (\ref{qnf}), one can observe that there indeed occur the minimum of the net force and the increase
following it (see Figure~\ref{fig:med}).
We point out that, in the dimensionless unit, both the zero temperature limit 
$\frac{l}{2 {\cal E}} \Delta F(0) = \frac{3}{4} N$ and the temperature $t$ at which 
the net force takes its minimum are of order $N$.

When we increase the temperature further, we find that the net force keeps
increasing, and actually proves to tend to infinity with a
square-root-of-temperature asymptotic behavior (see Figure 7).
This temperature dependence can be derived as follows.
By expanding $N_n^\pm$ in terms of $ q^\pm := e^{- \alpha^\pm} $ as
  \begin{equation}
  N_n^\pm = \frac{ q^\pm e^{- e_n^\pm / t} }{ 1 - q^\pm e^{- e_n^\pm / t} }
  = \sum_{k = 1}^{\infty} (q^\pm)^k e^{- k e^\pm_n / t} ,
   \label{aad} \end{equation}
we have 
  \begin{equation}
N = \sum\limits_{n = 1}^{\infty} N_n^\pm = \sum\limits_{k = 1}^{\infty}
(q^\pm)^k \sum\limits_{n = 1}^{\infty} e^{- k e_n^\pm / t}
= \sum\limits_{k = 1}^{\infty} (q^\pm)^k \left[ - \frac{\sigma^\pm}{2} +
\frac{1}{2} \sum\limits_{n = - \infty}^{\infty} e^{- k e^\pm_n / t} \right]
  \label{aae} \end{equation}
with $\sigma^+ = 0$ and $\sigma^- = 1$. Applying now the Poisson
summation formula,
\begin{equation}
 \sum_{n = - \infty}^{\infty} y(n) = \sum_{m = - \infty}^{\infty}
\int_{-\infty}^{\infty} ds\, y(s) e^{ 2 \pi i m s },
  \label{poisson} 
\end{equation}
we obtain
  \begin{equation}
N = \sum\limits_{k = 1}^{\infty} (q^\pm)^k \left[ - \frac{\sigma^\pm}{2}
+ \sqrt{ \frac{\pi t}{4 k} } \sum\limits_{m = - \infty}^{\infty}
(\mp 1)^m e^{ - \frac{\pi^2 t}{k} m^2 } \right] .
  \label{aag} \end{equation}
Similarly, for the forces $F^\pm$ [see (\ref{nfp})], we find
  \begin{equation}
F^\pm = \frac{2 {\cal E}}{l} \sum\limits_{k = 1}^{\infty} (q^\pm)^k
\sqrt{ \frac{\pi t^3}{16 k^3} } \sum\limits_{m = - \infty}^{\infty}
(\mp)^m \left( 1 - \frac{2 \pi^2 t}{k} m^2 \right)
e^{ - \frac{\pi^2 t}{k} m^2 } .
  \label{aah} \end{equation}
{}For the high-temperature asymptotic behavior ($N^\pm_1 \to 0 
\Rightarrow q^\pm \to 0 $), it suffices to consider only the first
two terms in the sums over $k$ in these sums, and within each term to keep
only the $m = 0$ term in the sums over $m$ (the $m \ne 0$ terms being
exponentially suppressed). Thus, from (\ref{aag}) we get

  \begin{equation}
q^\pm = 2 N / (\pi t)^{1/2} + 2 N ( \sigma^\pm - \sqrt{2} N ) / ( \pi t)
+ {\cal O} ( t^{-3/2} ),
  \label{aai} \end{equation}
and, correspondingly, the net force,
  \begin{equation}
\Delta F = \frac{{\cal E} N}{l} \left( \frac{t}{\pi} \right)^{1/2}
+ {\cal O} ( t^{0} ) ,
  \label{aaj} \end{equation}
which is the promised square-root-of-temperature asymptotic behavior.
We can see in Figure 7 how this asymptotic behavior is
reached at high temperatures.

The result that the net force does not tend to zero (nor
to a nonzero constant) seems unusual when contrasted to the naive
expectation that such quantum effects coming from the distinct
boundary conditions should vanish at high temperatures where the
classical picture would be available. However, this surprising
feature can be understood by the fact that, contrary to most
quantum systems, one dimensional wells have such energy spectra
that the level spacing is not decreasing but increasing for higher
energy levels (which is actually valid not only for boxes with
Dirichlet and/or Neumann boundary conditions but for all other
wells as well~\cite{FTC}). In other words, quantum wells can be
distinguished by their high-temperature behavior, too.

\begin{figure}[t] 
\begin{center}
\includegraphics[width=0.5\linewidth]{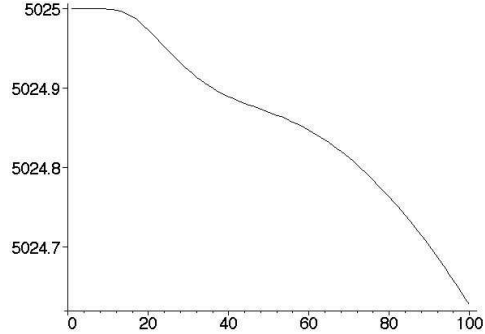}
\caption{The dimensionless net force 
$\frac{l}{2 {\cal E}} \Delta F(t)$ 
in the fermionic case, for $N = 100$, in the low-temperature
region.}
\label{fig:fer} 
\end{center} 
\end{figure}

One can replace the bosonic particles with fermions, and
consider the same problem as above, too. 
The net force is found to exhibit a
qualitatively similar temperature dependence as in the bosonic case, with
two main differences. One of them is that the $t = 0$ value of the net
force and the temperature where the net force takes its minimum are
proportional to $N^2$ rather than to $N$ as seen in the bosonic case.  
The other is that one
slight ``step" can be observed at low temperatures, where the net force
starts to decrease from its $t = 0$ value (see Figure~\ref{fig:fer}).
Otherwise the fermionic case is similar to the bosonic one, and, for
example, the high-temperature asymptotics proves to be the same
square-root-of-temperature one~\cite{FTtwo}.

\section{Discussions}

We have discussed in this paper some of the interesting physical phenomena that can arise on a line if there is a controllable point singularity either in the form of a dot or a wall.  It is remarkable that putting just a singular point on a line
allows for such variety of phenomena --- duality, anholonomy, supersymmetry, caustics anomaly and the emergence of pressure --- which are found usually in more involved
systems such as gauge field theory.  In addition, the scale anomaly ({\it i.e.} the breakdown of the classical scale symmetry at the quantum level) which we did not mention in this paper can also be seen in the system with a generic singularity.  This is implied by the presence of 
the scale parameters $L(\pm \theta)$, which are missing in the classical description.  
It is, therefore, safe to say
that the crucial element for those quantum phenomena to occur is not in the complexity of the system nor
in the infinity of the degrees of freedoms of the system as often assumed.  Rather, these are allowed because the quantum description of a system requires more information (and hence more parameters to be fixed) than the classical description does, and that once the extra parameters are chosen, some of the properties that hold classically may no longer hold, causing the anomalies at the quantum level.  The extra parameters in the case of a singularity on a line are given by the group $U(2)$, whose global structure is then used to yield the anholonomy effects, for example.  

Putting the singularity in many particle systems offers an interesting possibility when considered in the context of statistical mechanics.  We have seen this in the simple, albeit not too realistic, example of the quantum force acting on a partition wall in a square potential well.  The force attains a minimum at a certain temperature, which is proportional to the particle number $N$ for the bosonic case or to its square $N^2$  for the fermionic case, before it diverges for $T \to \infty$.  For the bosonic case
with $N \sim 100$, the minimum will be seen in a room temperature if the size of the system is about a few hundred nanometers, whereas for the fermionic case the same can be seen even with larger systems of the size of one micron.  In view of the rapid progress of nano-technology in recent years, it may not be entirely unreasonable to expect that some of these effects described here can be observed in laboratory in the near future.

\section*{Acknowledgements}

We thank our collaborators,
T.~Cheon, H.~Miyazaki and T.~Uchino 
for their valuable contributions and helpful discussions.
This work has been supported in part by 
the Grant-in-Aid for Scientific
Research on Priority Areas (No.~13135206) by
the Japanese
Ministry of
Education, Science, Sports and Culture.

\end{document}